\documentstyle[12pt]{article}
\newcommand{\beq}{\begin{equation}}
\newcommand{\eeq}{\end{equation}}
\def\lsim{\mathrel{\rlap{\lower3pt\hbox{\hskip0pt$\sim$}}
    \raise1pt\hbox{$<$}}}         

\def\gsim{\mathrel{\rlap{\lower4pt\hbox{\hskip1pt$\sim$}}
    \raise1pt\hbox{$>$}}}         

\begin{document}
\begin{titlepage}
\renewcommand{\thefootnote}{\fnsymbol{footnote}}

\begin{center} \Large
{\bf Theoretical Physics Institute}\\
{\bf University of Minnesota}
\end{center}
\begin{flushright}
TPI-MINN-94/42-T\\
UMN-TH-1323-94\\
hep-ph/9501222\\
31 December, 1994
\end{flushright}
\vspace{.3cm}
\begin{center} \Large
{\bf Determining $\alpha_s$ from Measurements at $Z$:
\\
How Nature Prompts us about  New Physics}
\end{center}
\vspace*{.3cm}
\begin{center} {\Large
M.  Shifman } \\
\vspace{0.4cm}
{\it  Theoretical Physics Institute, Univ. of Minnesota,
Minneapolis, MN 55455}
\\

\vspace*{0.2cm}

E-mail address: SHIFMAN@VX.CIS.UMN.EDU
\\

\vspace*{1cm}

{\Large{\bf Abstract}}
\end{center}

\vspace*{.2cm}

The value of $\alpha_s (M_Z)$ emerging from the so called global fits
based mainly on the data at the $Z$ peak (and assuming the standard
model)
is three standard deviations higher
than  the one stemming from the low-energy phenomenology. The
corresponding value of $\Lambda_{\rm QCD}$ is very large, $\sim$
500 MeV,
and is incompatible with crucial features of QCD. If persists, the
discrepancy
should be interpreted  as due to  contributions to the
$Z$-quark-antiquark
vertices which go beyond the standard model.

\end{titlepage}

\newpage

The statement that precision measurements of electroweak physics
at the $Z$
peak
give no evidence whatsoever of new physics lying beyond the
standard model
(SM)
is becoming  common place now.  Moreover, the values of the SM
parameters
extracted from analysis of the LEP and SLD data are used as
canonical. This
refers, in particular, to the strong  coupling constant. The so called
global fits
which assume validity of the standard model and are based on a
large set of
data (``high-energy data") yield  values of $\alpha_s (M_Z)$ in the
$\overline{MS}$ scheme
which cluster around 0.125 \cite{Vyso,Lang}), with the error bars
0.005
\cite{F1}. The corresponding value of $\Lambda_{\rm QCD}$
is about 500 MeV \cite{F2}. These numbers, accepted as the most
exact results
for the strong coupling constant existing at present, propagate
further into a
stream of papers, published in the last year or two, devoted to
various aspects of QCD.

The question arises whether Quantum Chromodynamics can  tolerate
these numbers. I will argue below that the answer is negative.  There
are two
reasons why I believe that $\alpha_s (M_Z)$ must be close to 0.11
and the
corresponding value of $\Lambda_{\rm QCD}$   close to 200 MeV
(or even smaller). A
more formal argument comes from consideration of traditional ``low-
energy"
data, the cleanest of which is the evolution of moments of the
structure
functions
in deep inelastic scattering (DIS). These measurements are  abundant
and
have high
statistics
(for a review see \cite{revdis}). Theoretical formulae for the moment
evolution are known  in the leading and next-to-leading logarithmic
approximations; if one considers the {\em Euclidean} domain of
$Q^2$ above $\sim$10 GeV$^2$ non-perturbative effects  play no
role in the Q$^2$ evolution of the moments provided that the
moments
considered are
not too high. A typical result for $\alpha_s$ emerging in this way
(scaled to the normalization point $\mu = M_Z$) is $0.113\pm 0.005$
\cite{Virch}.

Other ``low-energy" analyses -- heavy quarkonia,
jets at PEP and PETRA and so on -- produce similar and even lower
values of
$\alpha_s$, with the only exception to be discussed in some detail
below. However, in determining the value of $\alpha_s$ from these
data one encounters serious problems: due to the essentially
Minkowskean
nature of the jet and quarkonium calculations precise estimates of
the role of  non-perturbative
effects are difficult and, hence, the corresponding
results for $\alpha_s$ are inconclusive. Moreover, in some instances
above
even the perturbative next-to-leading corrections are not calculated
so far. It should be noted that lattice calculations (e.g.
Ref. \cite{lattice}) also produce $\alpha_s (M_Z) =0.115$
or less; being a lattice outsider I do not know how
reliable these lattice results are.

The second, less formal,  argument in favor of $\Lambda_{\rm
QCD}\lsim
200$ MeV
is the success of the operator product expansion (OPE) \cite{Wils}
in an extremely wide range of applications to different QCD problems.
Although less formal at the moment, this argument seems more
convincing
to
me. The analysis is based on the {\em Euclidean} expansions (QCD
sum rules
\cite{Shif1}) ensuring, thus, the best possible control over
non-perturbative contributions. Although Wilsonian OPE is valid in
any
consistent field theory, practically successful numerical predictions
become
possible only because of the existence of a {\em window} in QCD --
a crucial phenomenon not completely understood theoretically
\cite{NSVZ}. The window is a Euclidean domain of momenta where in
certain correlation functions perturbative corrections turn out
to be numerically smaller than non-perturbative ones. In typical
instances
after the Borel transformation the window stretches down to
$\sim 0.7$ GeV.  This fact ensures a very fast transition from
essentially
perturbative regime to essentially non-perturbative one. An
interpretation of the
window phenomenon emerges if one assumes that $b$, the first
coefficient
in the Gell-Mann-Low function, is a large parameter (numerically
large),
all perturbative corrections are suppressed by powers of $1/b$,
while the non-perturbative corrections do not contain this
parameter.
The window phenomenon would  be impossible if
$\Lambda_{\rm QCD} \gsim 500$ MeV. It should be stressed
that the difference between  $\alpha_s (M_Z) =0.125$ and
$\alpha_s (M_Z) =0.11$ (i.e. $\Lambda_{\rm QCD}\approx 500$
MeV {\em versus} 200 MeV) is not merely quantitative but, rather,
qualitative. In the first case one expects a rather slow and gradual
transition
from the perturbative regime to the non-perturbative one, while in
the
second case
the non-perturbative (power) corrections blow up in the domain
where perturbation theory still seems convergent.

Precise measurements of $\Lambda_{\rm QCD}$ from the QCD sum
rules
can be fully formalized. As a matter of fact, the old work \cite{Eide}
 in this direction is seminal. It yields $\Lambda_{\rm QCD}<210$
MeV
at the one-loop level.
 Later analyses along these
lines also exist. Further efforts, both theoretical and experimental,
are needed in order to measure $\Lambda_{\rm QCD}$ in this way
at the level of accuracy desirable today. But even the achieved level
of accuracy  rules out $\Lambda_{\rm QCD} \gsim 500$ MeV.

A discrepancy between the low-energy expectations for $\alpha_s$
and the fits at $Z$ alerted some theorists a few years ago (see below).
Their arguments were largely overshadowed later by the assertion,
worked out in a series of  interesting and stimulating papers
\cite{Pich},
that a precise low-energy determination of $\alpha_s$ from $\tau$ is
possible,
$\alpha_s (M_\tau) =0.33\pm 0.03$. This result implies, in turn, that
 $\alpha_s (M_Z) =0.120\pm 0.003$, in  accord with the value
measured at
the $Z$ peak. Ref. \cite{Pich} presents the state of the art in
estimating
all known sources of non-perturbative corrections. The leading
corrections
come from the gluon and four-quark condensates and turn out to be
negligibly
small. This calculation, quite correct by itself,
gives rise to a new doctrine  -- a very  precise determination of
$\alpha_s (M_Z)$ from essentially perturbative formula for
$\Gamma (\tau \rightarrow\mbox{hadrons})$ is possible  --
and we are witnessing
now how this doctrine  is gradually becoming  generally accepted
in the
community.

The problem with all approaches of this type is that it can not be
formulated
as a completely Euclidean analysis. One has to deal with
 Minkowskean spectral densities integrated with some weights over
some
{\em finite} energy range.
If in the genuinely Euclidean calculation one can reliably judge the
accuracy achieved by considering the retained correction terms,
an estimate of the accuracy for Minkowskean averages (integrals
over a finite energy range) based on
individual
condensate terms
is grossly misleading. To see that this is indeed the case it is
sufficient to
consider a model spectral density suggested in the last section of Ref.
\cite{Shif2}. This spectral density corresponds to an infinite series of
equidistant poles, with one and the same residue, and it may be
relevant
in the large $N_c$ limit in the channel with one heavy quark and a
massless
antiquark,
$$
\Pi_{\rm model} (E ) = -\psi (-E ) +\frac{1}{E}=
-\sum_{n=1}^\infty\frac{1}{E-n} +\mbox{Const}\, ,
$$
where $\psi$ is the logarithmic derivative of the $\Gamma $
function.
In order to mimic the standard QCD routine in the treatment of the
spectral density we proceed as
follows. One considers $\Pi (E )$ at Euclidean (negative) values of
$E$, expands in $1/E$ and then takes the imaginary part of the
expansion,
\beq
\Pi (E ) = -\ln (-E)+\frac{1}{2E} + \sum_{n>0} \frac{(-1)^{n-1}B_n}{2n}
\frac{1}{E^{2n}} \, ,
\eeq
\beq
{\rm Im} \, \Pi (E ) = \pi -\frac{\pi}{2}\delta (E) +\pi\sum_{n>0}
\frac{(-1)^{n-1} B_n}{(2n)!}\delta^{(2n-1)} (E)
\label{im}
\eeq
where $B_n$ stand for the Bernoulli numbers.
The constant term in ${\rm Im}\, \Pi (E )$ is an analog of the
``perturbative" term;
the rest is due to ``non-perturbative" power corrections.
Being extremely simple, the model contradicts  no general
requirements.

It is not difficult to check that the Borel transform of this function,
$\hat B\Pi
(\epsilon )$,
considered at Euclidean (negative) values of the Borel parameter
$\epsilon$,
possesses the property we expect from the OPE-based analysis;
namely, the exact result differs from the truncated series by a
quantity
of order of the last power term kept. At the same time the integrals
$$
\int_0^{E_0} E^n dE \, {\rm Im}\, \Pi
$$
calculated from the expansion (\ref{im}) differ from the exact values
by a large amount,
$\sim 1/E_0$, in spite of the fact that the power series for each given
moment
consists here of a finite number of terms. Thus, in the first moment
($n=0$),
if the first power term is retained, the second one and all others are
zero,
and one would expect the absolute accuracy following the line
of reasoning of Ref. \cite{Pich}.

Of course, the model considered is relevant only in the limit
$N_c=\infty$.
For $N_c=3$ a natural broadening of the resonances smears the
spectral
density and improves the accuracy of the Minkowskean calculations
done with
the truncated series. Still, the strength of the non-perturbative
effects
in the spectral density is represented not only by individual power
terms
of low dimensions but, also, by the asymptotic behavior of the power
terms of
high orders. The corresponding contribution is exponential in energy,
\beq
\Delta {\rm Im}\, \Pi \propto {\rm e}^{-CE} \sim \exp\{-C{\rm
e}^{C'/\alpha_s (E)}\}
\label{exp}
\eeq
 and,
therefore, is {\em not seen} in the truncated OPE expansions.
Equation
(\ref{exp})
gives an estimate of a deviation from duality which shows up
when one descends from the asymptotically high to lower energies.
$C$
and $C'$ in  are constants, see Ref. \cite{Shif2} and the  forthcoming
publication \cite{SUV} where the  issue will
be discussed in more detail.
At the moment no reliable {\em purely theoretical} method exists
that would allow one to find the constant $C$. In other words,
theoretical
estimates of non-perturbative contributions
in the  Minkowskean quantities of the type of the total
hadronic
width of $\tau$ or the $e^+e^-$ annihilation cross section at a given
energy
are rather vague,  and they definitely do not have such a great
accuracy as is required today in the problem of $\alpha_s$.
 One has to invoke phenomenological
information, and this reverses the problem. If we accept that
$\Lambda_{\rm QCD}$ lies in the vicinity of 200 MeV we have to
conclude
that about 20\% of the pre-asymptotic term in the hadronic $\tau$
width comes from non-perturbative effects (violations of duality) so
that
actually
$\alpha_s(M_\tau )\approx 0.27$. At the moment this conclusion
must be considered as perfectly legitimate. (By the preasymptotic
term
I mean $R_\tau - 3$ where $R_\tau$ is defined in \cite{Pich}.)
Similar and even larger violation of duality was shown \cite{BDS}
to take place in the inclusive semileptonic decays of the $D$ mesons
which are very close in mass to $\tau$'s.

What is usually done in the conference talks and review papers to
lull the public opinion is averaging of two groups of data
-- low-energy and high-energy values of $\alpha_s(M_Z)$.  Then the
world average usually quoted is $0.117\pm 0.005$; it lies only
$\sim 1.5$ standard deviations from either of them, and the
contradiction is
hidden under the rug.

If the value of $\alpha_s (M_Z)\approx 0.11$, as it stems from the
low-energy
data, what is the way out? The most placid solution of the problem
would be reversing the  trend of the $Z$ peak experiments. Only
three  years
ago the corresponding global fits used to yield numbers for the
strong  coupling
constant which did not contradict the above value. Since then the
result was  steadily increasing.

To bring the value of the strong coupling constant
in line with the low-energy considerations one has to diminish
the experimental number for the hadronic width of the $Z$ by
$\sim $ 7 MeV. Surprisingly, this 7 MeV is the excess of the
hidden beauty produced at $Z$, compared to the SM expectations,
detected recently \cite{Vyso,Lang}. As was noted in Refs.
\cite{Hold,Lang2},
if one allows for new physics in the $Zb\bar b$ vertex to
take care of this 7 MeV excess in $\Gamma (Z\rightarrow b\bar b)$
one then solves the $\alpha_s$ puzzle too.

If no systematic bias is found (and experts say that this scenario is
very unlikely)
we are forced to look for physical explanations of the discrepancy.
The SM global fits can be altered if there is a contribution due to new
physics.
As a matter of fact one of the explanations has been already
proposed --
light gluinos \cite{Clav}. Gluinos with masses of order of a few GeV
change the rate of running of the strong coupling (it becomes slower),
so that both numbers, the ``low-energy" $\alpha_s$ and the
``high-energy"
one become compatible with each other and compatible with the
estimate
$\Lambda_{\rm QCD}\approx 200$ MeV. Many theorists, however,
are
reluctant to accept this scenario because of certain specific problems
associated
with the
light gluinos. The niche for their existence is nearly closed
experimentally.

It seems more appealing to assume that new heavy particles (with
mass
$\gsim 100$ GeV ) generate, through loops,  a correction to the
$Z$-quark-antiquark vertices, enhancing the hadronic decays of the
$Z$.
In order to convert $\alpha_s (M_Z) =0.125$
into $\alpha_s (M_Z) =0.11$ it is sufficient to ensure the
enhancement of
the hadronic width by $\sim $ 0.4\%. The first idea that comes to
one's mind
is the fourth generation \cite{NVys}.
Then the
$Z$ boson has a well-defined
axial coupling to the doublet of quarks belonging to the fourth
generation;
let us denote them by $T$ and $B$. Through the
$T$ and $B$ loops the $Z$ boson proceeds into a pair of
gluons
which are then coupled to light quarks, $u,d,s,c,b$. (One of the gluon
propagators is contracted into a point due to the $Z$-boson quantum
numbers). Interference of this graph with the tree
$Z$-quark-antiquark
vertices produces a correction to the hadronic decays of $Z$
proportional
to $\ln m_{T}/m_{B}$, so that the sign of the effect can
be
adjusted at will. The elegance of this mechanism becomes obvious
if one takes into account the fact that the two-gluon intermediate
state is weak
isosinglet, so that the interference in
the $u\bar
u+d\bar d$ channel cancels.
The same
happens in the $s\bar s +c\bar c$ channel. The only surviving
correction is in the $b\bar b$ channel ($t$ is too heavy to appear
in the intermediate state and cancel it). Thus, the fourth generation
can
naturally enhance the $Z$ decays into $b\bar b$, the  mode
where the current experimental data are known to disagree with
theoretical expectations
at the $2\sigma$ level \cite{Vyso,Lang}. If one
could adjust the ratio $\ln m_{T}/m_{B}$
in such a way as to totally erase the disagreement in the $b\bar b$
then this
$b\bar b$
enhancement would be sufficient to simultaneously solve the
$\alpha_s$
problem.
Experts say, however \cite{NoVy,NORVY}, that a large ratio
$ m_{T}/m_{B}$ is ruled out by consistency of the SM
radiative
corrections to the masses and polarization operators of the $Z$ and
$W$ bosons. Stretching all numbers to their extremes I found
that the fourth generation can be responsible for at most 1.5 MeV in
$\Gamma (Z\rightarrow b\bar b)$, instead of
the desired 7.

Leaving the fourth generation aside we can turn to superpartners.
The light gluino scenario has been already mentioned. The heavy
(virtual)
gluino effects have been also discussed in the literature. The
correction is
generated by the $Z$ coupling to squarks which then exchange a
gluino
and convert into quarks. Both squarks and gluino
are assumed to lie in the 100 GeV ballpark. This mechanism was
studied in
\cite{Hagi},
and later, even in more detail, in \cite{Djuo}. The squarks/gluino
effect
in the $b\bar b$ channel is specifically addressed in the works
\cite{Boul}. As it follows from these calculations, the sign of the
gluino contribution is correct (i.e. it produces an enhancement) and,
moreover, the effect can reach the desired 0.4\% in the $Z$ hadronic
width
provided that the gluino/squark masses are on the light side of the
allowed mass domain. Thus, superpartners in loops  can,
in principle, solve both difficulties simultaneously.

It should be noted, though, that if the superpartners are responsible
for bringing  $\alpha_s(M_Z)$ down to 0.11 the possibility of
a straightforward Grand Unification within the Minimal
Supersymmetric Standard Model (MSSM) is ruled out.  Indeed,
the simplest version of Grand Unification, with the squark and
gluino masses in the 100 GeV ballpark, implies \cite{KKRW}
that $\alpha_s(M_Z)= 0.125$ or larger.
This seems to be an exciting observation
defying the standard boring great desert scenarios \cite{FL}.

Of course, one can say that the $\alpha_s (M_Z)$ problem --
0.125 versus 0.11 controversy -- is only a $3\sigma$ effect.  Being
translated
in the language of $\Lambda_{\rm QCD}$ the difference becomes
quite drastic.
Moreover, having $\Lambda_{\rm QCD}$ in the ballpark of 200 MeV
is crucial for consistency of a very large number of QCD-based
calculations
in the low-energy domain known to produce successful predictions
which, seemingly, will not  survive if $\Lambda_{\rm QCD}\gsim
500$ MeV.
In any case, the question is definitely ripe enough for an intensive
public
debate. The data on $\alpha_s (M_Z)$ from the $Z$ peak, as they
exist now, at
the very least
must be taken as a clear hint that new physics is around the corner.
Whether it
is supersymmetry or something else \cite{Hold} is hardly possible to
decide at
the
moment. The supersymmetric explanation is good since (i)
it goes through without twisting arms, i.e. it can naturally enhance
the total hadronic and $b\bar b$ widths of the $Z$ by the desired
amount without spoiling the rest of the well-fit picture with
the electroweak radiative corrections (see e.g. \cite{Alta}); (ii)
supersymmetry
is
anyway a popular element of the present-day theory as the only
available mechanism which might explain the lightness of the Higgs
particles
in a natural way (assuming, of course, that relatively light Higgs
particles do exist). A minimal lesson one has to draw is important for
QCD practitioners. At the moment it seems reasonable to abstain from
using
$\alpha_s (M_Z) = 0.125$ (and the corresponding value of
$\Lambda_{\rm QCD}$) as the best  measurement of these key QCD
parameters. If so, the problem of the semileptonic branching ratio
deficit in
the $B$ mesons \cite{BBSV}, which nearly disappeared \cite{Ball}
after the corresponding theoretical formulae were evaluated with the
large
$\alpha_s$, resurfaces again. It would be interesting to check
whether  penguins generated by superpartners can ensure sufficient
enhancement of the non-leptonic modes of $B$.
Work in this direction has already begun, with quite
encouraging results \cite{kaga}. The squark penguins can give rise
to the chromomagnetic operators of the type $\bar s
\sigma_{\mu\nu} G_{\mu\nu} b$ with the coefficients less
suppressed compared to the standard model expectations
\cite{VZS}. Then the  $b\rightarrow s$ + gluon transition
can be responsible for, say,  $20\%$ of the hadronic width of the $B$
meson, thus eliminating any difficulties with ${\rm Br}_{sl}(B)$
\cite{kaga}. Simultaneously, the expectation value of the
charm multiplicity in the $B$ decays goes down, which is also
welcome.  Further analysis is needed to check the overall
consistency. It is necessary to verify, for instance, that the
$b\rightarrow s\gamma$ rate is not enhanced beyond what is
acceptable.  Similar squark penguins can play a role in the $\Delta I =
1/2$ rule in the strange particle decays.

Another question to  be
considered is as follows. If the violation of duality at $\tau$ is
at the level of 20\% of the pre-asymptotic term, what is to be
expected
in the inclusive $B$ decays?

All questions discussed above have been repeatedly considered in
the literature previously -- different aspects in separate publications.
I combine them together.
The only element which I  add is my deep conviction that
$\Lambda_{\rm QCD}$ can not be larger than
$\sim$ 200 MeV. The measurements of $\alpha_s$ at the Z pole
must be interpreted as a direct indication on new physics. Since
convictions are very hard to formalize this letter should be viewed as
an open invitation to further discussions among experts.

\vspace{0.5cm}

Illuminating  discussions with K. Hagiwara,  H. Ohnishi, L. Roszkowski,
M. Virchaux  and especially P.  Langacker,
V. Novikov and
M. Vysotsky are gratefully acknowledged. I would like to thank
B. Holdom, P. Langacker and M. Vysotsky for pointing out to me
Refs. \cite{Hold,Lang2,NORVY,Boul}. This work was supported
in part by
DOE under the grant number
 DE-FG02-94ER40823.


\begin{thebibliography}{99}
\bibitem{F1}
Ref. \cite{Lang} quotes even a higher value,
$\alpha_s(M_Z)=0.127\pm0.005$.
The errors are believed to be dominated by theoretical uncertainties.
The
most precise calculation is done for the total hadronic width
$\Gamma (Z\rightarrow\mbox{hadrons})$. The error quoted above,
$\Delta
\alpha_s (M_Z)
=\pm 0.005$, is essentially determined by analyzing
$\Gamma (Z\rightarrow\mbox{hadrons})$. If these data are
excluded,
all other measurements at $Z$ (event shape, jets,...) yield
$\alpha_s(M_Z) = 0.120\pm 0.008$ where I quote the average value
obtained
in Ref. \cite{Ohni}.
\bibitem{F2}
At the present level of accuracy it is important to specify what
particular
definition of the QCD scale parameter is accepted. Most commonly
used is an
artificial parameter called $\Lambda_{\overline{\rm MS}}$, see Ref.
\cite{revdis} for an excellent review. The fact that
$\Lambda_{\overline{\rm
MS}}$ does not directly measure the scale where $\alpha_s$ becomes
unity
and is, thus, quite inconvenient, is clearly seen from the $N_f$
dependence of $\Lambda_{\overline{\rm MS}}$ ($N_f$ is the
number
of flavors).  To make contact with other papers easier I will continue
using $\Lambda_{\overline{\rm MS}}$, with a regret. Results
extracted from
deep inelastic scattering (DIS) are usually expressed in terms of
$\Lambda_{\overline{\rm MS}}^{(4)}$ while those referring to
the $Z$ peak are expressed in terms of
$\Lambda_{\overline{\rm MS}}^{(5)}$. The numerical relation
between these
two parameters depends on where exactly one matches the running
laws
referring to $N_f=4$ and $N_f=5$. The corresponding uncertainty
in $\alpha_s (M_Z)$ is $\sim 0.001$. The scale parameter which is
most
relevant for the genuine low-energy hadronic physics is
$\Lambda_{\overline{\rm MS}}^{(3)}$. The absolute majority of
authors
avoid giving this parameter in fearing of introducing further
uncertainties
associated with
matching at the charmed quark threshold. Therefore, below I will
quote the
values of $\Lambda_{\overline{\rm MS}}^{(4)}$, denoting this
parameter,
by definition, by $\Lambda_{\rm QCD}$. The reader should
remember
that $\Lambda_{\overline{\rm MS}}^{(4)} \approx (3/2)
\Lambda_{\overline{\rm MS}}^{(5)}$ and, moreover,
$\Lambda_{\overline{\rm MS}}^{(3)}$ is larger than
$\Lambda_{\overline{\rm MS}}^{(4)}$. The value of the ratio
$\Lambda_{\overline{\rm MS}}^{(3)}/\Lambda_{\overline{\rm
MS}}^{(4)}$
is rather sensitive to the point of matching.
\bibitem{FL}
By the simplest version of Grand Unification I mean the one which
ignores subtleties due to possible GUT-scale thresholds and
high-dimension
(nonrenormalizable) operators. Including these effects one can,
perhaps, descend somewhat below $\alpha_s(M_Z)$=0.125. How
much below --
this question requires further investigation.
\end{thebibliography}
\end{document}